\documentclass{elsart}

\usepackage[dvips]{graphicx}
\usepackage{amsmath,amssymb}
\usepackage{bm}
\begin{document}

\begin{frontmatter}



\title{Cosmic rays from thermal sources}

\author[a]{G. Wilk\corauthref{cor1}}
\address[a]{The Andrzej So{\l}tan
Institute for Nuclear Studies, Ho\.{z}a 69, 00-681 Warsaw, Poland}
\ead{wilk@fuw.edu.pl} \corauth[cor1]{Corresponding author}
\author[b]{Z. W\l odarczyk}
\address[b]{Institute of Physics,
\'Swi\c{e}tokrzyska Academy, \'Swi\c{e}tokrzyska 15, 25-406
Kielce, Poland} \ead{wlod@pu.kielce.pl}

\begin{abstract}
Energy spectrum of cosmic rays (CR) exhibits power-like behavior
with very characteristic "knee" structure. We consider a
generalized statistical model for the production process of cosmic
rays which accounts for such behavior in a natural way either by
assuming existence of temperature fluctuations in the source of
CR, or by assuming specific temperature distribution of CR
sources. Both possibilities yield the so called Tsallis statistics
and lead to the power-like distribution. We argue that the "knee"
structure arises as result of abrupt change of fluctuations in the
source of CR. Its possible origin is briefly discussed.
\end{abstract}


\end{frontmatter}

\section{Introduction}
\label{Int}

The origin of the characteristic features of the energy spectrum
of cosmic rays (CR), which has power-like behavior with "knee"
structure, remains matter of hot debate (for survey of models
proposed to explain the origin of CR see \cite{CR-origin}). It
could reflect different regimes of diffusive propagation of CR in
the Galaxy, but it could also be due to some property of
acceleration processes within the source of the CR itself. In this
second case, a crucial question is whether the sources of CR below
the "knee" can also accelerate particles to much higher energies,
so that a single population of astrophysical objects can explain
the smooth spectrum of cosmic rays, as observed over many orders
of magnitude in energy. We address this problem using a
generalized statistical model specially adapted to this end.
However, our work will concentrate more on the physics of CR than
on the generalized statistics, providing therefore some physical
explanations to ideas presented already in \cite{TAB} and
\cite{CB}. In particular, we shall argue that the observed "knee"
structure of the CR energy spectrum arises as result of some
abrupt change of fluctuation pattern in the source of CR.

\section{Nonextensive statistics and results}

We shall start with some basic information on nonextensive
statistical mechanics as introduced by Tsallis \cite{T}, which
have been already successfully applied to a variety of complex
physical systems, including CR where, among others, the energy
spectrum of cosmic rays have been analyzed from a nonextensive
point of view \cite{TAB,CB}. The idea is to maximize the more
general entropy measure than the usual Boltzman-Gibbs-Shanon (BGS)
entropy, the one which depends on a new additional parameter $q$
and which leads to generalized version of the statistical
mechanics. If we optimize, under appropriate constrains, the BGS
entropy,
\begin{equation}
S = - \int dE\, P(E)\, \ln P(E) , \label{eq:BGS}
\end{equation}
we obtain the equilibrium distribution in the usual form of
exponential distribution,
\begin{equation}
P(E) = \frac{1}{T}\, \exp \left( - \frac{E}{T} \right).
\label{eq:BGeq}
\end{equation}
This distribution can alternatively be obtained as the solution of
simple differential equation:
\begin{equation}
\frac{dP(E)}{dE}\,  = \, - \frac{P(E)}{T} . \label{eq:difeqBG}
\end{equation}
A more general formalism proposed in \cite{T} (and sometimes
referred to as nonextensive statistical mechanics) is based on the
generalized entropy,
\begin{equation}
S_q \, =\, - \frac{\int dE\, P^q(E)\, -\, 1}{q-1}\, \, .
\label{eq:Sq}
\end{equation}
Its maximization under appropriate constrains yields a
characteristic power-like distribution (which sometimes is also
called $q$-exponential distribution, $\exp_q(\dots)$):
\begin{equation}
P_q(E)\, =\, \frac{2-q}{T}\, \exp_q\left( - \frac{E}{T}\right)\,
=\, \frac{2-q}{T}\, \left[ 1\, -\, (1-q) \frac{E}{T}
\right]^{\frac{1}{1-q}}. \label{eq:P_q}
\end{equation}
For $q \rightarrow 1 $ one recovers the usual exponential
distribution (\ref{eq:BGeq}). This equilibrium distribution can
alternatively be obtained by solving the following differential
equation
\begin{equation}
\frac{dP(E)}{dE} \, =\, - \frac{P^q(E)}{T}. \label{eq:difeqT}
\end{equation}
Precisely this equation has been used in \cite{TAB} to describe
the flux $\Phi(E)$ of cosmic rays. However, the values of the
temperature obtained there seem to be uncomfortably high.

On the other hand there is growing evidence that nonextensive
formalism applies most often to nonequilibrium systems with a
stationary state that posses strong fluctuations of the inverse
temperature parameter $\beta = 1/T$ \cite{WW,B}. In fact,
fluctuating $\beta$ according to gamma distribution with variance
$Var(\beta)$ results in a power like distribution (\ref{eq:P_q})
with nonextensivity parameter $q$ being given by the strength of
these fluctuations,
\begin{equation}
q-1 \, =\, \frac{Var(\beta)}{\langle \beta\rangle^2},
\label{eq:varbeta}
\end{equation}
distribution (\ref{eq:BGeq}) is just its limiting case when
$q\rightarrow 1$. This observation was used in \cite{CB} to
describe the flux $\Phi(E)$. Although the results were reasonably
good the estimated temperature $T\sim 170$ MeV (comparable with
the so called Hagedorn temperature known from description of
hadronization processes) seems to be, again, overestimated. This
was because author insists on description of the whole range of
energy spectrum including its very low energy part which, in our
opinion, is governed mainly by the geomagnetic cut-off and should
be considered separately.

\subsection{Energy spectrum}

For relativistic particles (where the rest mass $m$  can be
neglected) the energy $E\sim p$ and the density of states of an
ideal gas in three dimensions is given by  $\omega (E) \propto
E^2$. The flux $\Phi(E)$ can be then obtained straightforwardly
from $P(E)$ and reads
\begin{equation}
\Phi(E)\, =\, N_0\, E^2\, P(E), \label{eq:flux}
\end{equation}
where  $N_0$ is normalization factor. For $E>>T$  we have power
spectrum $\Phi(E) \propto E^{-\gamma}$ with the slope parameter
$\gamma $ which in terms of parameter $q$ introduced above is
$\gamma = (3-2q)/(q-1)$. In the case of CR this spectrum  has the
shape of broken power law $E^{-\gamma}$ which changes pattern in
the region named as "knee" with  the slope $\gamma_1 \simeq 2.7$
at energies below $\sim 10^{15}$ eV and $\gamma_2 \simeq 3.1$ at
energies between the "knee" and the highest measurable energies
$E\sim 10^{18}$ eV. In the language of the nonextensivity
parameters it would mean that $q_1 = 1.213$ before and $q_2 =
1.196$ after the "knee", i.e., one can argue then that at the
"knee" one witnesses the change of fluctuation pattern.

\subsection{Temperature fluctuations}

As mentioned above the special role in converting exponential
distribution to its $q$-exponential counterpart play fluctuations
of the scale parameter provided in the form of gamma function.
There are at least two scenarios leading to gamma distribution in
$\beta$ mentioned above (here $\mu^{-1} = \beta_0(q-1)$ and
$\nu^{-1} = q-1$):
\begin{equation}
f(\beta)\, =\, \frac{\mu}{\Gamma(\nu)}\left( \mu \beta\right)^{\nu
-1}\, \exp\left( - \beta \mu\right), \label{eq:Gammadistr}
\end{equation}
\begin{itemize}
\item[$(a)$] temperature distribution of sources; \item[$(b)$]
temperature fluctuations in small parts of a source.
\end{itemize}

In what concerns the first possibility notice that gamma
distribution (\ref{eq:Gammadistr}) is the most probable outcome of
the maximalization of Shannon information entropy (\ref{eq:BGS})
under constraints that $\int f(\beta) d\beta =1$, $\langle
\beta\rangle = \beta_0$ and (because distribution we are looking
for is one sided, i.e., defined only for $\beta > 0$) that
$\langle \ln (\beta)\rangle = \ln \left(\nu \beta_0\right)$.
However, this possibility is rather unlikely because in this case
one expects that there is some cut temperature $T_{cut}$ such that
$T < T_{cut}$, which would result in very characteristic rapid
break in the CR energy spectrum, not observed in experiment.

To illustrate the second possibility let us suppose that one has
thermodynamic system, different small parts of which have locally
different temperatures, i.e., its temperature understand in the
usual way fluctuates. Let $\xi(t)$ describes the stochastic
changes of temperature in time and let it be defined by the white
Gaussian noise ($\langle \xi(t)\rangle = 0$ and $\langle \xi (t)
\xi (t+\Delta t)\rangle  = 2 D \delta(\Delta t)$). The inevitable
exchange of heat which takes place between the selected regions of
our system leads ultimately to the equilibration of temperature.
As we have advocated in \cite{WW}, the corresponding process of
the heat conductance leads eventually to the gamma distribution
$f(\beta)$ (\ref{eq:Gammadistr})  with variance (\ref{eq:varbeta})
related to the specific heat capacity $C_V$ of the material
composing this system by
\begin{equation}
q = 1 + \frac{1}{C_V} . \label{eq:C_V}
\end{equation}
The change of fluctuation pattern in the "knee" region mentioned
before would therefore correspond in this case to abrupt change in
the heat capacity of the order of $C_2/C_1 = 1.09$.

\begin{figure}[h]
\vspace{-6mm}
\begin{center}
\includegraphics [width=2.5in]{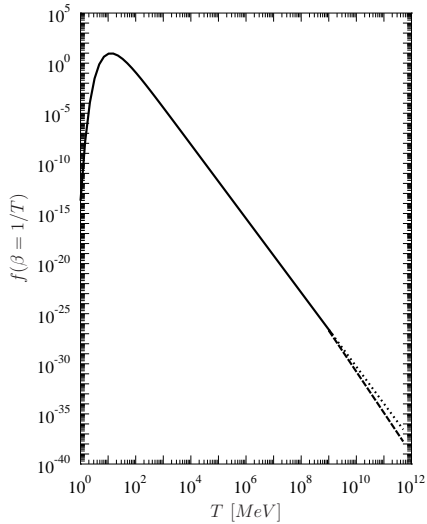}
\end{center}
\vspace{-0.9cm}
 \caption{Example of the possible modification of gamma distribution
needed to describe fluctuation of temperature which could lead to
description of the "knee region" of the CR spectra. At $T =
10^{9}$ MeV parameter $q$ changes from $q = 1.214$ (it would
result in dotted line) to $q = 1.2$ (in this case curve changes
slope and continues as full line). Such break represents some
abrupt change in the fluctuation patter, which we attempt to
explain here.} \label{fig0}
\end{figure}

\subsection{Heat capacity}

Can one expect something of this kind to happen in the
astrophysical environment of the CR? In what follows we shall
argue that, indeed, one can. Let us first notice that subject of
temperature fluctuations in astrophysics is much-discussed problem
nowadays. Its  effect on the temperatures empirically derived from
the spectroscopic observation was first investigated in \cite{P}
whereas in \cite{13a,13b,13c,13d} it was shown that temperature
fluctuations in photoionized nebulea have great importance to all
abundance determinations in such objects. It means that discussion
of the heat capacity or, equivalently, the behavior of parameter
$q$ defining the energy spectrum, is fully justified.

Let us concentrate therefore on the problem of heat capacity of
astrophysical objects, in particular in neutron stars. In such
objects one observes the following feature. The total specific
heat of their crust, $C$, is the sum of contributions from the
relativistic degenerate electrons, from the ions and from
degenerate neutrons. In the temperature that can be reached in the
crust of an acreating neutron star (which is of the order of $T
\sim5\cdot 10^8$ K and is below the Debaye temperature $T_D \sim
5\cdot 10^9$ K) we have $C_{ion} < C_e < C_n$. When the
temperature drops below the critical value $T = T_C$ the neutrons
become superfluid and their heat capacity $C^{sf}_n$ increases
\cite{HC1,HC2},
\begin{eqnarray}
\frac{C^{sf}_n}{C_n} \simeq 3.15\frac{T_C}{T} \exp\left( -
1.76\frac{T_c}{T}\right)\cdot \left[2.5 - 1.66
\left(\frac{T}{T_C}\right) + 3.68 \left(\frac{T}{T_C}\right)^2
\right] .\label{eq:C/C}
\end{eqnarray}
At  $T \sim 0.7 T_C$ we have $C^{sf}_n \sim 1.1 C_n$ what
corresponds to the changes of spectral index by $\Delta \gamma
\sim 0.5$. To summarize: one witnesses here the abrupt change in
the heat capacity at some temperature.

\vspace{1cm}
\begin{figure}[h]
\begin{center}
\includegraphics [width=7.cm]{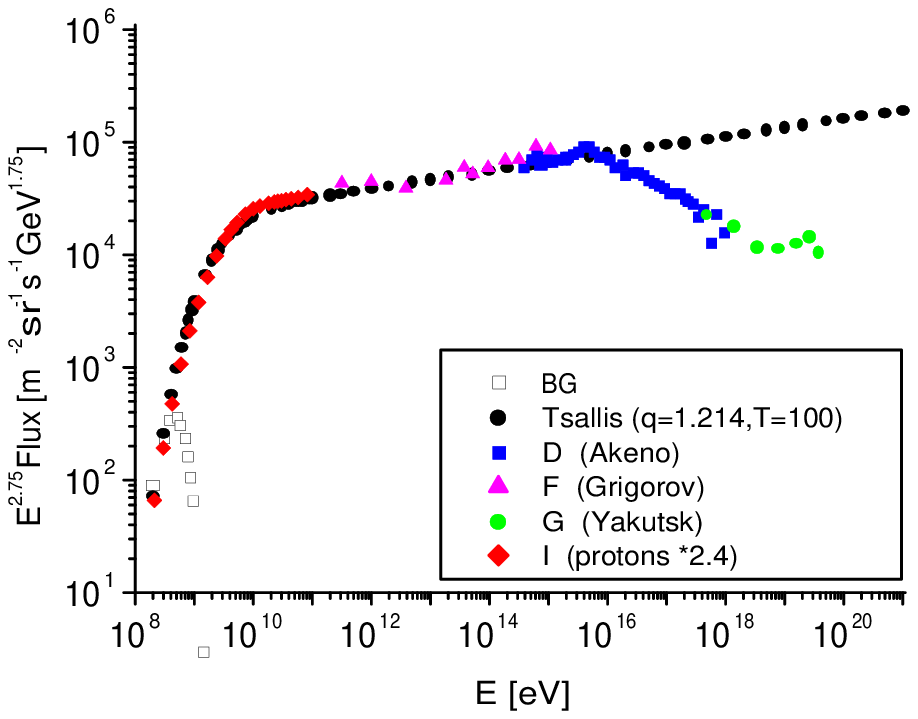}\hspace{5mm}
\includegraphics [width=7.cm]{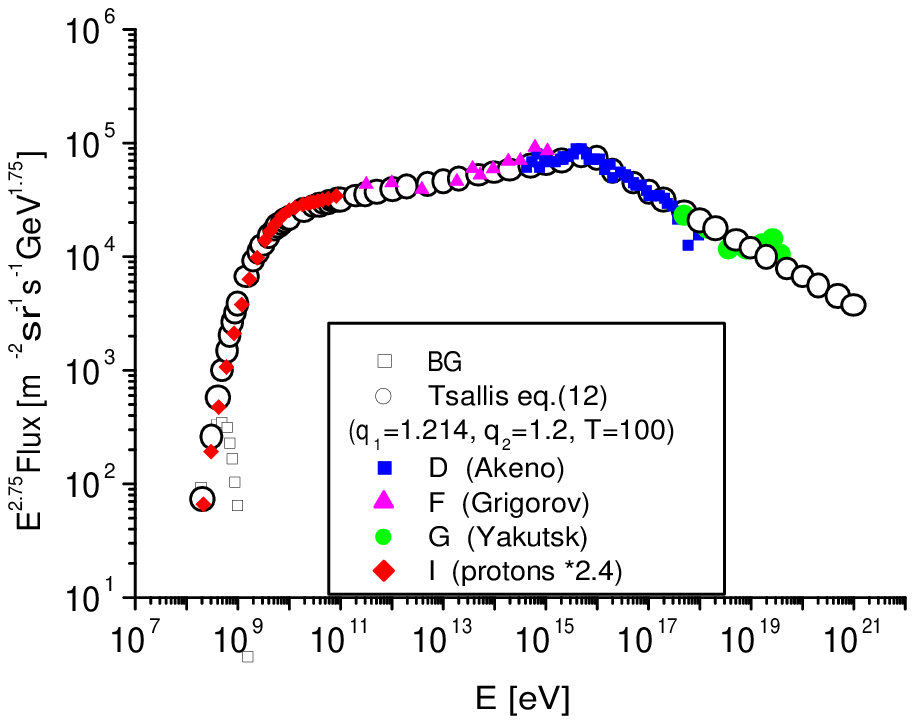}
\vspace{-5mm} \caption{CR energy spectra fitted by simple Tsallis
distribution (left panel) and distribution using specially adopted
fluctuation function example of which is presented in Fig.
\ref{fig0} (right panel). Notice the total inadequacy of simple
exponential (Boltzman, denoted by BG here) distribution in
description of these spectra.} \label{fig1}
\end{center}
\end{figure}

Suppose now that we take seriously the conjecture expressed by eq.
(\ref{eq:C_V}) that $C_V$ is directly connected with the parameter
$q$. It means then that, the usual fluctuation pattern given by
gamma distribution (\ref{eq:Gammadistr}) should be replaced by it
slightly modified version shown in Fig. \ref{fig0}, which is
characterized by two nonextensivity parameters, $q_1$ and $q_2$.
This change is assumed to be abrupt and taking place at some
temperature $T_{cut}$ and it differs our proposition from what was
proposed in \cite{TAB}, which from our point of view, corresponds
to some form of composition of two (suitably normalized) such
gamma functions with different $q$ each. Following our proposition
one obtains the following flux of CR:
\begin{eqnarray}
\Phi(E) = N_0\, E^2 \cdot  \left[P_{q_1}(E) - \alpha_1 (E)
P_{q_1}(E) + \alpha_2(E) P_{q_2}(E)\right] ,\label{eq:fluxchange}
\end{eqnarray}
where $P_q(E)$   is given by eq. (\ref{eq:P_q}) and $$ \alpha_i =
\Gamma \left[\frac{1}{q_i-1}, \, \frac{1-\left(1-q_i\right)
E/T}{\left(q_i-1\right) T_{cut}/T}\right]/\Gamma
\left(\frac{1}{q_i -1}\right).$$ Our results are presented in Fig.
\ref{fig1}. The "knee" region is reproduced very well, however,
the price to be paid is the need of suitable choice of energy at
which fluctuation pattern changes (cf. Fig.
\ref{fig0})\footnote{Two remarks are in order at this point.
First, the nucle\-on superfluidity was predicted already in
\cite{Migdal} and today pulsar glitches provide strong
observational support for this hypothesis \cite{NS}. Nucleon
superfluidity arises from the formation of Cooper pairs od
fermions (actually in \cite{qSUP} also quark superfluidity from
cooling neutron stars were investigated). Continuous formation and
breaking of the Cooper pairs takes place slightly below $T=T_C$
(critical temperature $T_C$ is in the order $10^9 - 10^{10}$ K).
The other, neutron stars are born extremely hot in supernova
explosions, with interior temperatures around $T\sim 10^{12}$  K.
Already within a day, the temperature in the cental region of the
neutron star will drop down to $\propto 10^9 - 10^{10}$ K and will
reach $10^7$ K in about $100$ years \cite{NeStar}. The first
measurements of the temperature of a neutron star interior (core
temperature of the Vela pulsar is $T \sim 10^8$ K, while the core
temperature of PSR B0659+14 and Geminga exceeds $2\cdot 10^8$ K)
allow to determine the critical temperature $T_C \sim 7.5\cdot
10^9$ K \cite{critT}.}.

\subsection{Acceleration}

As one can see in Fig. \ref{fig1} we have obtained agreement with
data for the whole range of CR energy spectrum but the price paid
for this is, again, apparently too high value of the temperature
parameter used, $T=100$ MeV. We have prone therefore to the same
kind of criticism as we have applied to previous attempts in this
field \cite{TAB,CB}. The possible way out of this dilemma is to
argue that the break in the original spectrum and connected with
it phase transition occurs actually at much slower energies and
that resultant spectrum is then accelerated to the observed
energies - for example by the magneto-hydrodynamical turbulence
and/or shock discontinuities (i.e., by the so called DSA
mechanism, cf. \cite{DSA}). The simplest version of this
mechanism, as discussed in \cite{FM}, implies that distribution
function after the shock, $f_{after}(p)$, is related to the
original distribution before the shock, $f_{before}(p)$ in the
following way:
\begin{equation}
f_{after}(p) = \frac{b}{p^b}\, \int_{p_{min}}^p dp'\, p'^{(b-1)}\,
f_{before}(p'), \label{eq:DSA}
\end{equation}
where $b=3r/(r-1)$. Here $p$ denotes the particle momentum,
$r=\rho_2/\rho_1$ describes the compression of densities across
the shock and $p_{min}$ denotes the minimal value of momenta. DSA
mechanism transforms a $\delta\left(p-p_0\right)$ spectrum of
relativistic particles in a power-like spectrum of the type $f(p)
\propto p^{-b}$. For example, if one has initial spectrum of the
form $\propto p^{-c}$ which encounters a shock with strength given
by $b$ then eq. (\ref{eq:DSA}) shows that:
\begin{itemize}
\item in the case of $b>c$ (which corresponds to the initial
spectrum being softer than it would result from a
$\delta$-function injected into the shock) one has $f_{after}
\propto p^{-c}$, i.e., the acceleration does not change the shape
of the spectrum; \item in the case of $b<c$ (i.e., for the steep
initial spectrum) one has $f_{after} \propto p^{-b}$, which
coincides with result of injection of a $\delta$-function into the
shock.
\end{itemize}
It can be then shown from eq. (\ref{eq:DSA}) that when an ensemble
of shocks is encountered, the shape of the spectrum should be
given by the strongest shock \cite{FM}.

That such scenario is {\it a priori} plausible in the case
considered here can be seen by the following argumentation. If the
energy growth is given by
\begin{equation}
\frac{\partial E}{\partial t} \, =\, a + bE, \label{eq:partial1}
\end{equation}
than from master equation
\begin{equation}
\frac{\partial P(E)}{\partial t} \, =\, - c\, P(E)
\label{eq:partial2}
\end{equation}
one gets the evolution equation
\begin{equation}
\frac{\partial P(E)}{\partial E} \, =\, -c P(E)\, \frac{\partial
t}{\partial E} \, =\, - \frac{ P(E)}{T(E)} \label{eq:partial3}
\end{equation}
in the form of eq.(\ref{eq:difeqBG}) with energy dependent
temperature parameter:
\begin{equation}
T(E)\, =\, \frac{a+bE}{c}\, =\, \frac{T_0 + (q-1)E}{q},
\label{eq:T(E)}
\end{equation}
where we have used: $c=T_0^{-1}$, $a=q^{-1}$  and
$b=(1-q^{-1})T_0^{-1}$. Energy dependent temperature parameter
$T(E)$ in this form immediately leads to the energy distribution
given by eq.(\ref{eq:P_q}). Notice that $q^{-1}$  plays here the
role of the weight with which we select the constant (thermal) and
linear (accelerating) terms in the equation describing the growth
of energy. It means therefore that Tsallis distribution preserves
its structure when subjected to the aforementioned acceleration
scheme. The problem, which remains to be solved is whether the
broken spectrum with the "knee" structure can be suitably
transformed in the same way. In particular, whether the "knee"
structure is preserved and how its position before the
acceleration process is related to that actually observed.

To summarize this part: stochastic mechanisms of acceleration of
CR particles (like acceleration on the fronts of shock waves or
Fermi acceleration in turbulent plasmas, both analogous in some
sense to Brownian motion) do not change the shape of the
production spectra, but, unfortunately, they are not particularly
effective, i.e., they do not lead to large increase of energy
\cite{Gaisser}. The increase of energy per one collision is of the
order $\Delta E/E \sim u^2/c^2$, what for the plasma velocity $u
\thickapprox 10^6 - 10^7$ cm/s gives $\Delta E/E \thickapprox
10^{-8}$ and leads to the mean relative increase of energy during
the time life of Galaxy ($t\thickapprox 3\cdot 10^{17}$ s) only by
factor $\langle \delta\rangle \thickapprox 3$. Fluctuation on the
steep spectrum of accelerated particles result in additional
increase of energy. Because of the multiplicative character of
acceleration we have log-normal distribution of variable $\delta$,
$P(\delta) d\ln \delta = \left(\sqrt{2\pi}\sigma \right)^{-1}
\exp\left[-\left( \ln \delta - \ln \langle \delta \rangle
\right)^2/2\sigma^2 \right] d\ln \delta$, what results in shift of
the spectrum of source on the energy scale by $\langle \delta
\rangle ^{(\gamma -1)/\gamma} \exp\left[(\gamma -
1)^2\sigma^2/(2\gamma)\right]$, where $\sigma^2$ is variation of
the distribution $P(\delta)$. For $\gamma \thickapprox \langle
\delta \rangle \thickapprox \sigma^2 \thickapprox 3$ we can obtain
only order of magnitude shift of the energy spectrum.

\section{Summary and conclusions}

Let us summaries arguments presented above. \begin{itemize} \item
The spectrum of cosmic rays has the shape of broken power law
$E^{-\gamma}$ , with the slope $\gamma_1 \simeq 2.7$ at energies
below $\sim 10^{15}$ eV and $\gamma_2 \simeq 3.1$ at energies
between the knee and $E\sim 10^{18}$ eV. This slopes correspond to
the nonextensivity parameters (taking into account that
$q=(3+\gamma)/(2+\gamma)$) $q_1 = 1.213$ and $q_2 = 1.196$,
respectively, i.e., to the change of heat capacity of the order of
$C_2/C_1 = 1.09$. \item Nonextensive statistics successfully
describes the smooth power-law spectrum and traces its origin back
to fluctuations of temperature, $f(\beta)$, being given by gamma
distribution. \item Out of two possible scenarios leading to such
distribution of inverse temperature we prefer the temperature
fluctuation in the source rather than the temperature distribution
of sources. The point is that in the second case some cut
temperature $T_{cut}$ is expected which would result in the rapid
break in the energy spectrum and which is not observed. The
temperature $T$ (not essential in the high energy region, $E>>T$)
seems to be of the order of MeV, i.e. of the order of the interior
stars temperature (if stars are born extremely hot in supernova
explosions, with interior temperatures around $T\sim 100$ MeV,
already within a day the temperature in the cental region of the
star will have dropped down to $\sim 0.1 - 1$ MeV and reach the
$1$ keV in about $100$ years). The critical temperature
(corresponding to the nucleon superfluidity) is $T_C \sim 0.1 - 1$
MeV. It means then that the origin of changes of the
nonextensivity parameters at temperature $T_{cut} \simeq 10^{15}$
eV $\simeq 10^{19}$ K is still open question. \item The
nonextensive formalism leads to production (injection) spectrum
and the acceleration processes ($dE/dt \sim E$, which does not
change the shape of power spectrum) and allows the shift of this
spectrum to high energies. The question of how to connect "knee"
position before and after such acceleration process remains,
however, still open.
\end{itemize}

\begin{figure}[h]
\begin{center}
\hspace{-2cm}
\includegraphics [width=11.5cm]{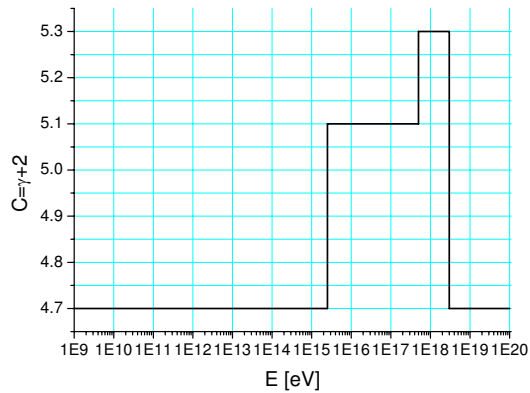}
\end{center}
\vspace{-9cm} \caption{Pattern of the energy dependence of the
heat capacity obtained from the CR energy spectrum in the vicinity
of the "knee" structure using nonextensive statistical approach.}
\label{fig2}
\end{figure}

Let us close with some very intriguing observation. Independently
of the discussion presented above one can notice that the measured
CR energy spectrum can be converted using arguments presented
above (eq. (\ref{eq:C_V}) connecting Tsallis parameter $q$ with
the heat capacity) into energy dependence of the heat capacity
$C$. The result is shown in Fig. \ref{fig2}. As one can see $C$
acts here as a kind of magnifying glass converting all subtle
structures of $\Phi(E)$ into much more pronounced and structured
bump. Its importance would parallel long-standing discussion of
the origin of the knee-like structure of the CR energy spectrum,
but exposed in much more dramatic and visible way. At the moment
we can only offer two examples of the possible explanation of this
feature. The first is that this effect is due to the change of the
effective number of degrees of freedom in the incoming projectiles
with energy. Assuming that most of CR consist of protons, which
are build from three quarks, one could speculate that each bumps
correspond to excitations from single proton to proton plus one
quark and two quark structure, after exciting all three quarks one
comes back to the original situation (much in the spirit of the
changes observed when ice becomes water and this becomes steam -
there also corresponding $C$'s show characteristic jump
\cite{Water}). \vspace{-1cm}
The other, perhaps more realistic explanation of Fig. \ref{fig2},
is to connect it with the behavior of heat capacity in Fermi
liquids. Following \cite{HC1} the proton heat capacity $C\sim
m^*/m$, i.e., it is proportional to the ratio of the effective
mass of the proton in the neutron fluid to the mass of the free
proton. In the case of a mixture of Fermi liquids the proton
effective mass $m^*$ is affected by interactions with neutrons and
other protons and is given by
\begin{equation}
\frac{m^*}{m} = 1 + \frac{1}{3}D_p\left[ f^{pp}_1 + \left(
\frac{k_{F_n}}{k_{F_p}}\right)^2 f_1^{pn}\right],
\label{eq:mstaroverm}
\end{equation}
where $D_p$ denotes the density of quasiparticle states at Fermi
surface given by wave vectors $k_{F_n}$ and $k_{F_p}$ for,
respectively, neutrons and protons whereas $f^{pp}_1$ and
$f^{pn}_1$ are Landau parameters \cite{BJK}. Fig. \ref{fig2} can
be then interpreted as showing changes of $C$ with energy in the
Fermi liquid. We start with the superfluid liquid with $C_1=4.7$
(here $m^*$ represents effective mass for $pp$ and $pn$
interactions),  when energy increases we stop to see nuclear
interactions and $C_2 = 5.1$ (with $m^*$ representing $pp$
interactions only), finally, for large $T$, one has Fermi gas with
$C_3 =5.3$ and, still further, the usual Fermi liquid\footnote{It
is worth to remember that fluctuations of temperature we are
talking about in this work refer to fluctuations in small region
$V$. For Fermi liquid the heat capacity expressed in units of
Boltzmann constant $k_B$ (i.e., for $k_B=1$) is of the order
$C\simeq 3\cdot 10^{35}$ cm$^{-3}$ \cite{YU}. Therefore, taking
values of $C$ estimated from the slope of the primary CR spectra
(cf. Fig. \ref{fig2}) one gets that size of the region of
fluctuations is $V \sim 10^4$ fm$^3$.}. Notice that
\begin{equation}
\frac{1}{3}D_p f_1^{pp} = \frac{C_2-C_3}{C_3}\qquad {\rm
and}\qquad \frac{1}{3}D_p\left(
\frac{k_{F_n}}{k_{F_p}}\right)^2f_1^{pn} = \frac{C_1-C_2}{C_3},
\label{eq:estimations}
\end{equation}
what results in the following relation between Landau parameters,
\begin{equation}
\left( \frac{k_{F_n}}{k_{F_p}}\right)^2 \frac{f_1^{pn}}{f_1^{pp}}
= \frac{C_1 - C_2}{C_2 - C_3} = 2. \label{eq:LP}
\end{equation}
In the case of a one-component Fermi liquid we have the well known
identity, $m^*/m = 1 + F^{pp}_1/3$, where $F_1^{pp} =
D_pf_1^{pp}$. From (\ref{eq:LP}) we can see that in two-component
Fermi liquid the quantity $1-m^*/m$ is $3$ times bigger (this is
because parameter $f_1$ which determines interaction between
quasiparticles is negative what results in smaller effective
mass). From properties of excited states in nuclear matter ($Pb$
and neighbor nuclei \cite{SZR}) $F_1^{np} = - 0.5 \pm 0.25$. If
$F_1^{nn} < F_1^{np} < F_1^{pp}$ and taking (after \cite{PH})
$F_1^{nn} - F_1^{pp} = -0.2$, we can estimate that for
neutron-star matter one has $m^*/m = 1 + F_1^{pp} \simeq 1 - 0.4
\pm 0.3 = 0.6 \pm 0.3$.

We end by saying that the above are just plausible examples, not
fully convincing explanation(s). It means then that this problem
deserves further scrutiny to be performed elsewhere.

\section*{Acknowledgment}

Partial support (GW) of the Ministry of Science and Higher
Education under contracts 1P03B02230 and  CERN/88/2006 is
acknowledged.

\end{document}